\pdfoutput=1
\documentclass[aps,prd,preprint,superscriptaddress,tightenlines,nofootinbib]{revtex4}

\long\def\symbolfootnote[#1]#2{\begingroup%
\def\thefootnote{\fnsymbol{footnote}}\footnote[#1]{#2}\endgroup}

\usepackage{amsfonts,amsmath,amsthm,amssymb}
\usepackage{enumitem}
\usepackage{mathrsfs}
\usepackage{bm}
\usepackage{color}
\usepackage{epsfig}

\newcommand{\PRE}[1]{{#1}}   

\usepackage[usenames,dvipsnames]{xcolor}
\definecolor{orange}{cmyk}{0,0.5,1,0}
\definecolor{rossoCP3}{cmyk}{0,.88,.77,.40}
\definecolor{graa}{rgb}{0.8,0.8,0.8}
\definecolor{blaa}{rgb}{0.2,0.2,0.6}

\begin{document}

\title{\PRE{\vspace*{0.3in}} \color{rossoCP3}{Supersymmetric sphaleron
    configurations \\ as the origin of the perplexing ANITA events
}}
\PRE{\vspace*{0.1in}} 

\author{\bf Luis A. Anchordoqui}

\affiliation{Department of Physics and Astronomy,\\  Lehman College, City University of
  New York, NY 10468, USA
\PRE{\vspace*{.05in}}
}

\affiliation{Department of Physics,\\
 Graduate Center, City University
  of New York,  NY 10016, USA
\PRE{\vspace*{.05in}}
}

\affiliation{Department of Astrophysics,\\
 American Museum of Natural History, NY
 10024, USA
\PRE{\vspace*{.05in}}
}

\author{\bf Ignatios Antoniadis}

\affiliation{LPTHE, Sorbonne Universit\'e, CNRS\\ 4 Place Jussieu, 75005 Paris, France
\PRE{\vspace*{.05in}}}

\affiliation{Albert Einstein Center, Institute for Theoretical Physics\\
University of Bern, Sidlerstrasse 5, CH-3012 Bern, Switzerland
\PRE{\vspace*{.05in}}}

\begin{abstract}\vskip 2mm
 \noindent The ANITA experiment has observed two air shower events
  with energy $\sim 500~{\rm PeV}$ emerging from the Earth with exit
  angles of $\sim 30^\circ$. We explain ANITA events as arising from
  neutrino-induced supersymmetric sphaleron transitions. These
  high-multiplicity configurations could contain a large number of
  long-lived supersymmetric fermions, which can traverse the Earth and
  decay in the atmosphere to initiate upward-pointing air showers at
  large angles above the horizon. We comment on the sensitivity of new
  generation LHC detectors, designed to searching for displaced decays of
  beyond standard model long-lived particles, to test our model.
  \end{abstract}

\maketitle

The $SU(3)_C \otimes SU(2)_L \otimes U(1)_Y$ standard model (SM) of
electroweak and strong interactions has recently endured intensive
scrutiny at the Large Hadron Collider (LHC) using a dataset
corresponding to an integrated luminosity of $63.9~{\rm fb}^{-1}$ of
2018 $pp$ collisions at center-of-mass energy $\sqrt{s} = 13~{\rm
  TeV}$, and it has proven once again to be a remarkable structure
that is consistent with all experimental results by tuning more or
less 19 free parameters. However, the Antarctic Impulsive Transient
Antenna (ANITA) experiment, designed to observe ultrahigh-energy
cosmic rays and neutrinos from outer space, has detected particles
that seemed to be blasting up from Earth instead of zooming down from
space, challenging SM
explanations~\cite{Gorham:2016zah,Gorham:2018ydl}. As a matter of
fact, several beyond standard SM physics models have been proposed to
accommodate ANITA
observations~\cite{Cherry:2018rxj,Anchordoqui:2018ucj,Huang:2018als,Dudas:2018npp,Connolly:2018ewv,Fox:2018syq,Collins:2018jpg,Chauhan:2018lnq},
but a convincing explanation is yet to see the light of day. In this
Letter, we entertain the possibility that ANITA events originate in a
supersymmetric sphaleron transition produced in the scattering of
extremely high-energy ($E_\nu \agt 10^{10.5}~{\rm GeV}$) cosmic
neutrinos with nucleons inside the Earth. Such a non-perturbative
process yield a high-multiplicity final state containing several
long-lived supersymmetric fermions, one of which would survive
propagation through the Earth crust before decaying into SM particles
to initiate an upward-pointing shower in the atmosphere, just below
the ANITA balloon.

The advantages of our interpretation of ANITA events over previous
supersymmetry (SUSY) models ~\cite{Fox:2018syq,Collins:2018jpg} go in two directions:
\begin{itemize}[noitemsep,topsep=0pt]
\item The ratio ${\rm BR} (\nu N \to {\rm SUSY})$ of the
  neutrino-nucleon cross section into SUSY particles over the total
  $\nu N$ cross section dominates over the branching ratio of charged
  current (CC) $\nu N$ interactions. Furthermore, the particle content of
  the final state in sphaleron-induced transitions could contain a
  large multiplicity of SUSY fermions. All of this is in sharp
  contrast with the production of SUSY pairs in perturbation theory,
  for which ${\rm BR} (\nu N \to {\rm SUSY}) \alt
  10^{-4}$~\cite{Albuquerque:2003mi,Albuquerque:2006am,Ahlers:2006pf,Ahlers:2007js,Ando:2007ds}.
\item The $\nu N$ scattering process requires a center-of-mass energy
  $\sqrt{s} \agt 245~{\rm TeV}$, thus probing $E_\nu \agt
  10^{10.5}~{\rm GeV}$. In this energy range a large
   flux of neutrinos is expected from the decay of cosmic
  strings~\cite{Berezinsky:2011cp}.  Moreover, in our model all three
  neutrino flavors would contribute to the ANITA signal.
\end{itemize}
We begin our discussion by highlighting the main characteristics of
ANITA events and after that we provide a phenomenological analysis of
data.

After three balloon flights, the ANITA experiment has detected two
perplexing upgoing showers with energies of ($600 \pm 400)~{\rm
  PeV}$~\cite{Gorham:2016zah} and
($560^{+300}_{-200}$)~PeV~\cite{Gorham:2018ydl}.\footnote{The trigger
  algorithm used for the second flight was not sensitive to this type of events~\cite{Gorham:2016zah}.} The energy estimates
are made under the assumption that the showers are initiated close to
the event's projected position on the ice. These estimates are lowered
significantly if the showers are initiated far above the ice. For
example, the energy of the second event is lowered by 30\% if the
shower is initiated four kilometers above the
ice~\cite{Gorham:2018ydl}. Note that even with the 30\% energy
reduction, the center-of-mass energy of the collisions initiating
these showers is beyond  $\sqrt{s}$ of the LHC beam.

In principle, ANITA events could originate in the atmospheric decay of
an upgoing $\tau$-lepton produced through a CC interaction of a
$\nu_\tau$ inside the Earth~\cite{Feng:2001ue}. However, the
relatively steep arrival angles of these events ($27.4^\circ$ and
$34.5^\circ$ above the horizon) create a tension with the SM
neutrino-nucleon interaction cross section. More concretely, the
second event implies a propagating chord distance through the Earth $=
2 R_\oplus \cos \theta_n \sim 7.2 \times 10^3~{\rm km}$, which
corresponds to $1.9 \times 10^4~{\rm km}$ water equivalent (w.e.)  and
a total of 18~SM interaction lengths at $E_\nu \sim 10^3~{\rm
  PeV}$~\cite{Dutta:2005yt}. Here, $R_\oplus$ is the radius of the
Earth and $\theta_n$ the nadir angle of the event. The first event
emerged at $\theta_n \simeq 62.6^\circ$ implying a chord through the
Earth of $5.9 \times 10^3~{\rm km}$, which corresponds to $1.5 \times
10^4~{\rm kmw.e.}$ for Earth's density profile~\cite{Gorham:2016zah}.
Because the energy deposited in a shower is roughly 80\% of the
incident neutrino energy, the cosmic neutrino energy range of interest
is $200 \alt E_\nu/{\rm PeV} \alt 1000$.  Taking the view that the
event distribution is maximized at $\theta_n = 60^\circ$, in our
calculations we will consider an average chord distance in traversing
the Earth of $\sim 6 \times 10^3~{\rm km}$.

Next, in line with our stated plan, we study the structural properties
of our model. In the mid-seventies 't Hooft pointed out that the SM
does not strictly conserve baryon and lepton
number~\cite{'tHooft:1976up,'tHooft:1976fv}.  Rather, non-trivial
fluctuations in $SU(2)$ gauge fields generate an energy barrier
interpolating between topologically distinct vacua.  An index theorem
describing the fermion level crossings in the presence of these
fluctuations reveals that neither baryon nor lepton number is
conserved during the transition, but only the combination $B-L.$
Inclusion of the Higgs field in the calculation modifies the original
instanton configuration~\cite{Klinkhamer:1984di}.  An important aspect
of this modification (called the ``sphaleron'') is that it provides an
explicit energy scale $E_{\rm sph} \sim M_W/\alpha_W \sim 9~{\rm TeV}$
for the height of the barrier, where $M_W$ is the mass of the charged
vector bosons $W^\pm$ and $\alpha_W \simeq 1/30$. When the energy
reach is much lower than $E_{\rm sph}$ the tunneling rate through the
barrier is exponentially suppressed $\Gamma_{\rm tunneling} \propto
e^{- 4\pi /\alpha_W} \sim e^{-164}$. However, the sphaleron barrier
can be overcome through thermal transitions at high temperatures,
providing an important input to any calculation of cosmological
baryogenesis~\cite{Kuzmin:1985mm,Fukugita:1986hr,Arnold:1987mh}. Indeed,
the rate over the barrier (thermal excitation) contains a Boltzmann
factor $\Gamma_{\rm thermal} \propto T^4 e^{-E_{\rm sph}/T}$, and
hence the rate becomes large as the temperature approaches $M_W$.

More speculatively, it has been suggested that the topological
transition could take place in two particle collisions at very high
energy~\cite{Aoyama:1986ej,Ringwald:1989ee,Espinosa:1989qn}.  The
anomalous electroweak contribution to the partonic process can be
written as \begin{equation} \hat\sigma_i(\hat s)= 5.3\times 10^3 \
  e^{-(4\pi/\alpha_W)\ F_W(\epsilon)}\  {\rm
    mb} \
  , \label{sigmahat} \end{equation} where the tunneling suppression
exponent $F_W(\epsilon)$ is usually refer to as the ``holy-grail
function'' and $\epsilon \equiv \sqrt{\hat s} /(4\pi M_W/\alpha_W)\simeq \sqrt{\hat s}/30~{\rm
  TeV}$~\cite{Khlebnikov:1990ue,Khoze:1990bm,Mueller:1991fa}. Altogether,
it is possible that at or above the sphaleron energy the cross section
could be of ${\cal O}(\rm mb)$~\cite{Ringwald:2003ns}.

The argument for strong damping of the anomalous cross section for
$\sqrt{\hat s}\gtrsim $~30~TeV was convincingly demonstrated in~\cite
{Bezrukov:2003er,Bezrukov:2003qm}, in the case that the classical
field providing the saddle point interpolation between initial and
final scattering states is dominated by spherically symmetric
configurations. This $O(3)$ symmetry allows the non-vacuum boundary
conditions to be fully included in extremizing the effective
action. In~\cite{Gould:1993hb} it was shown that a {\em sufficient}
condition for the $O(3)$ dominance is that the interpolating field
takes the form of a chain of ``lumps'' which are well-separated, so
that the each lump lies well into the exponentially damped region of
its nearest neighbors. However, we are not aware of any reason that
such lumped interpolating fields should dominate the effective
action. It is thus of interest to explore the other extreme, in which
non-spherically symmetric contributions dominate the effective action
(and let experiment rather than
theory~\cite{Tye:2015tva,Funakubo:2016xgd,Tye:2017hfv} be the
arbiter).  Thus far, the searches for instanton-induced processes in
LHC data have shown no evidence for excesses of high-multiplicity
final states above the predicted
background~\cite{Ellis:2016ast,Sirunyan:2018xwt,Ringwald:2018gpv}.

Of particular interest here would be an enhancement of the
$\nu N$ cross section over the perturbative SM estimates, say
by an order of magnitude, for $E_\nu \agt 10^{10.5}~{\rm GeV}$. To
get an estimate of this cross section we first note that for the
simple sphaleron configuration $s$-wave unitarity is violated for
$\sqrt{\hat s}> 4\pi M_W /\alpha_W\sim 36~{\rm
  TeV}$~\cite{Ringwald:2003ns}. If for $\sqrt{\hat s} > 36~{\rm TeV}$
we saturate unitarity in each partial wave, then this yields a
geometric parton cross section $\pi R^2$, where $R$ is some average
size of the classical configuration. As a fiducial value we take the
core size of the Manton-Klinkhamer sphaleron, $R \sim 10^{-2}~{\rm
  fm}$. In this simplistic model, the $\nu N$ cross section is found
to be
\begin{equation}
\sigma_{\nu N}^{\rm black \, disk} (E_\nu) 
 = \pi R^2\ \int_{x_{\rm min}}^1 \sum_{\rm partons} f(x)\ dx \,,
\end{equation}
where $x_{\rm min} = \hat s_{\rm min}/ s = (36)^2/ 2m_NE_\nu \simeq
0.065$, where $m_N$ is the mass of an isoscalar nucleon, $N \equiv (n
+ p)/2$, in the renormalization group-improved parton model. In the
region $0.065 < x < 3 \, (0.065)$ the parton distribution function for
the up and down quarks is well approximated by $f\simeq 0.5/x$, so the
expression for the cross section becomes
\begin{equation}
\sigma_{\nu N}^{\rm black \, disk} (E_\nu) 
\simeq \pi R^2\,\, (0.5)\,\, (\ln 3)\,\,  (2/2) 
 \simeq  1.5\times 10^{-30} {\rm cm}^2  \,,
\end{equation}
where the last factor of 2/2 takes into account the (mostly) 2
contributing quarks $(u,d)$ in this range of $x,$ and the condition
that only the left-handed ones contribute to the scattering. This is
about 80 times the SM cross section. Of course this calculation is
very approximate and the cross section can easily be smaller by a
factor of 10 (e.g., if $R$ is 1/3 of the fiducial value used). The
sphaleron production cross section derived
``professionally''~\cite{Ellis:2016dgb} is consistent with our
back-of-the-envelope estimate, and shows an enhancement of the $\nu N$
cross section over the perturbative SM estimates by about an order of
magnitude in the energy range $E_\nu \agt 10^{10.5}~{\rm
  GeV}$. Previous estimates pointed to even larger cross section
enhancements above perturbative SM prediction~\cite{Han:2004kq,Ahlers:2005zy}. In
our calculations we will adopt the estimate of~\cite{Ellis:2016dgb}.

A point worth noting at this juncture is that the energy for the
height of the barrier in SUSY models is also about
10~TeV~\cite{Moreno:1996zm}, and consequently the expected production
rate of supersymmetric sphaleron configurations is comparable to the
SM one~\cite{Cerdeno:2018dqk}. Most importantly, the decay BR
increases if the final state contains a large number of SUSY
fermions~\cite{Cerdeno:2018dqk}.  To develop our program in the
simplest way, we will work within a construct with gauge mediated SUSY
breaking, in which the gravitino $\psi_{3/2}$ is the lightest
supersymmetric particle (LSP) and the next-to-lightest supersymmetric
(NLSP) is a long-lived bino $\tilde B$~\cite{Giudice:1998bp}. Note
that for $M_{\tilde B} \sim 700~{\rm GeV}$~\cite{Aaboud:2018jiw}, NLSPs could be copiously
produced through instanton-induced processes at $\sqrt{\hat s} \agt
50~{\rm TeV}$ (see Fig.~3 in~\cite{Cerdeno:2018dqk}), and could
propagate inside the Earth without suffering catastrophic energy
losses from electromagnetic interactions.  The bino decays into a gravitino and a gauge boson (i.e., photon or $Z$-boson) with
Planck-suppressed partial widths,
\begin{eqnarray}
  \Gamma (\tilde{B}\rightarrow\psi_{3/2}\gamma) &=&
  \frac{\cos^2 \theta_{\rm W}}{48\pi M_{\rm Pl}^2}
  \frac{M_{\tilde{B}}^5}{m_{3/2}^2}
  (1 - x_{3/2}^2)^3 (1 + 3 x_{3/2}^2),
  \\
  \Gamma (\tilde{B}\rightarrow\psi_{3/2} Z) &=&
  \frac{\sin^2 \theta_{\rm W} \beta_{\tilde{B}\rightarrow\psi_{3/2} Z}}
  {48\pi M_{\rm Pl}^2}
  \frac{M_{\tilde{B}}^5}{m_{3/2}^2}
  \Big[
    (1 - x_{3/2}^2)^2 (1 + 3 x_{3/2}^2)  - x_Z^2
    \nonumber \\ & \times &
     \left\{
      3 + x_{3/2}^3 (-12 + x_{3/2}) + x_Z^4 
      - x_Z^2 (3 - x_{3/2}^2)
    \right\} 
  \Big],
\end{eqnarray}
where $M_{\rm Pl} \sim 10^{19}~{\rm GeV}$, $\sin^2 \theta_W \approx 0.23$, $x_{3/2}\equiv
m_{3/2}/M_{\tilde{B}}$, and $x_Z\equiv M_Z/M_{\tilde{B}}$, and where 
\begin{equation}
  \beta_{\tilde{B}\rightarrow\psi_{3/2} Z} \equiv
  \left[ 
    1 - 2 (x_{3/2}^2 + x_Z^2) 
    + (x_{3/2}^2 - x_Z^2)^2 
  \right]^{1/2},
\end{equation}
for $M_{\tilde{B}}>m_{3/2}+M_Z$, and
$\beta_{\tilde{B}\rightarrow\psi_{3/2} Z}=0$ otherwise~\cite{Feng:2004mt}.  For
$M_{\tilde{B}}>m_{3/2}+M_Z$, the total decay width is well
approximated by
\begin{equation}
  \tau_{\tilde{B}}^{-1} \simeq 
  \Gamma (\tilde{B}\rightarrow\psi_{3/2}\gamma) 
  + \Gamma (\tilde{B}\rightarrow\psi_{3/2} Z),
\end{equation}
and the NLSP lifetime is estimated to be
\begin{equation}
\tau_{\tilde B} \sim   5 \times 10^{14} \ \frac{m_{3/2}^2}{M_{\tilde
    B}^5}~{\rm s} \, ,
\label{lifetime}
\end{equation}
when masses are given in GeV~\cite{Covi:2009bk}.

Before proceeding, we pause to discuss existing limits from searches
of long-lived neutral particles at the Tevatron and at the LHC. The
CDF Collaboration searched for long-lived particles which decay to
$Z$-bosons by looking for $Z \to e^+e^-$ decays with displaced
vertices and excluded proper decay lengths $c \tau < 20~{\rm cm}$
for masses $< 110~{\rm GeV}$~\cite{Abe:1998ee}. Searches by D0
Collaboration exclude long-lived neutral particles of comparable
lifetimes and masses~\cite{Abazov:2006as,Abazov:2009ik}. The CMS
Collaboration has searched for long-lived neutralinos decaying into a
photon and an invisible particle, excluding $c \tau < 50~{\rm cm}$ for
masses $< 220~{\rm GeV}$~\cite{Chatrchyan:2012jwg}. The
  ATLAS Collaboration searched for high-mass long-lived particles that
  decay within the inner detector to give displaced dilepton vertices
  excluding $c \tau < 100~{\rm cm}$~\cite{Aad:2015rba}. ATLAS has also
  searched for very low mass ($< 10~{\rm GeV}$) long-lived particles
  by considering pairs of highly collimated
  leptons~\cite{Aad:2014yea}, with sensitivity to $c\tau \alt 20~{\rm
    cm}$. The most restrictive constraints on the lifetime of a
  long-lived particle come from a search by the ATLAS Collaboration
  for final states with displaced dimuon vertices in collisions at
  $\sqrt{s} = 13~{\rm TeV}$~\cite{Aaboud:2018jbr}.  Proper decay
  lengths $c \tau < 14~{\rm m}$ are excluded for SUSY models in which the
  lightest neutralino is the NLSP, with a relatively long lifetime due
  to its weak coupling to the LSP-gravitino. The lifetime limits are
  determined for very light gravitino mass and a
  neutralino mass of 700~GeV. Altogether, we can
  remain consistent with LHC bounds requiring $\tau_{\tilde B} \sim 
  44~{\rm ns}$ for $M_{\tilde B} \sim 700~{\rm GeV}$.  
Substituting the bino lifetime in (\ref{lifetime}) we obtain $m_{3/2} \sim 122~{\rm keV}$.

SUSY models with a gravitino LSP are also constrained by a variety of
cosmological observations. Of relevance to our analysis: {\it (i)}~if
$\tau_{\tilde B} \sim 44~{\rm ns}$, NLSP decay does not perturb light
element abundances which are synthesized during Big Bang
nucleosynthesis~\cite{Kawasaki:2008qe,Kawasaki:2017bqm}; {\it (ii)}~if $m_{3/2} \sim 122~{\rm keV}$, the
relic density of gravitinos can be accommodated to match observations
with choice of parameters~\cite{Fujii:2003iw,CahillRowley:2012cb}.

It takes a proper time of order $4.5 M_W^{-1}$ until the sphaleron
radiation shows free-field behavior~\cite{Hellmund:1991ub}. For
neutrino-induced sphaleron transitions, this radiation will be emitted
in a cone with half-opening angle $\delta \phi \sim {\cal O}
(1/\gamma)$, where $\gamma$ is the Lorentz factor. Taking fiducial
values $E_\nu \sim 10^{10.5}~{\rm GeV}$ and $\sqrt{\hat s} \sim
50~{\rm TeV}$, one can have an order of magnitude estimate $\gamma
\sim 6 \times 10^5$. All in all, the bino decay length in the lab
frame is $\gamma c \tau \sim 8 \times 10^3~{\rm km}$. This means that
for emerging angles $\theta_n \sim 60^\circ$, a long-lived bino could
survive the trip through the Earth. Note also that the boosted bino
would have an energy $E_{\tilde B} \sim 420~{\rm PeV}$, and after
decay roughly half of its energy will be deposited in the air shower.
These order of magnitude estimates are in good agreement with the
energy and opening angle distributions shown in Fig.~4
of~\cite{Ellis:2016dgb}.

 Given an isotropic $\nu + \bar \nu$ flux, the number of
  binos that emerge from the Earth is proportional to an ``effective
  solid angle'' $\Omega_{\rm eff} \equiv \int d \theta_n d \phi \cos
  \theta_n P(\theta_n,\phi,X)$, where $P(\theta_n,\phi;X)$ is the
  probability for a neutrino with incident nadir angle $\theta_n$ and
  azimuthal angle $\phi$ to emerge as a detectable $\tilde
  B$~\cite{Kusenko:2001gj,Anchordoqui:2001cg}.  $P(\theta,\phi,X)$ is
  a rather complicated function of various unkown (model dependent)
  parameters $X$. However, we can provide  a rough estimate of the
event rates if we adopt the exposure calculations
of~\cite{Fox:2018syq}, which suggest a total ANITA exposure for
sub-EeV emergent cosmic rays of $2.7~{\rm km}^2 \, {\rm sr} \, {\rm
  yr}$, for the two flights together.  It is noteworthy that this
exposure is orders of magnitude larger than the exposure for
$\tau$-neutrinos reported by the ANITA
Collaboration~\cite{Romero-Wolf:2018zxt}. This is because
$\tau$-neutrinos which do not arrive at very large nadir angles are
mostly blocked by the Earth. Observation of 2 events at ANITA would
require an integrated neutrino flux $\Phi_\nu ({E_\nu > 10^{10.5}~{\rm
    GeV}}) \sim 10^{-17.7}~({\rm cm}^2 \, {\rm s} \, {\rm
  sr})^{-1}$. Interestingly, at $E_\nu \sim 10^{10.5}~{\rm GeV}$, the
ANITA experiment sets the most restrictive upper limit on the energy
weighted cosmic neutrino flux; namely, $E_\nu \Phi_\nu (E_\nu) \alt
10^{-17.5}~({\rm cm}^2 \, {\rm s} \, {\rm sr})^{-1}$ at 90\%
CL~\cite{Allison:2018cxu,Aartsen:2018vtx}.  Note that neutrino-induced
sphaleron transitions with non negligible (missing) energy carried
away by long-lived SUSY fermions would relax limits on the neutrino
flux at extreme energies.  We end with two comments on the neutrino
flux. On the one hand, the required flux level to accommodate ANITA
events may be exceptionally high by astronomical
standards~\cite{Waxman:1998yy}.  On the other hand, for some model
parameters, such a flux of extremely high-energy ($E_\nu \agt
10^{10.5}~{\rm GeV}$) neutrinos is consistent with predictions from
decay of cosmic strings~\cite{Berezinsky:2011cp}. The decay of cosmic
strings also produces extremely high-energy photons and electrons that
interact with the cosmic microwave background and extra galactic
background light, producing an electromagnetic cascade, whose energy
density is constrained by measurements of the diffuse $\gamma$-ray
background~\cite{Ackermann:2014usa}. A point worth noting at this
juncture is that the fluxes of $\gamma$-rays and neutrinos expected from
the decay of cosmic strings are consistent with existing
observations~\cite{Anchordoqui:2018qom}. Moreover, experiments are
being designed to search for the neutrino signals of cosmic strings; 
e.g., the Lunar Orbital Radio Detector (LORD) that will fly aboard the Luna-Resurs
Orbiter space mission~\cite{Ryabov:2016fac}.

In summary, we have provided an interpretation of ANITA events in
terms of neutrino-induced supersymmetric sphaleron transitions.  These
high-multiplicity $B+L$ violating transitions may contain a large
number of long-lived SUSY fermions, which can traverse the Earth and
decay in the atmosphere to initiate an upward-pointing shower just
below the ANITA balloon. As a proof of concept, we have framed our
discussion in the context of a gauge-mediated breaking scheme, but this
model spans only a small region of the SUSY parameter space that can
accommodate ANITA events. Indeed, our interpretation of these perplexing events 
can be encapsulated in the product of three factors:
\begin{itemize}[noitemsep,topsep=0pt]
\item the differential flux of incident neutrinos,
\item the ratio of the $\nu N$ cross section into SUSY particles
over the total $\nu N$ cross section,
\item the lifetime of the SUSY fermion.  
\end{itemize}
Note these three factors are actually generic to a broad class of
models in which the messenger of ANITA events does not live inside the
Earth neither originate at cosmological distances. New generation LHC
experiments dedicated to searching for long-lived particles (such as
the ForwArd Search ExpeRiment
(FASER)~\cite{Feng:2017uoz,Ariga:2018uku}, the MAssive Timing
Hodoscope for Ultra Stable neutraL pArticles
(MATHUSLA)~\cite{Chou:2016lxi,Curtin:2018mvb}, and the Compact
Detector for Exotics at LHCb (CODEX-b)~\cite{Gligorov:2017nwh}) will
provide an important test both of the last two factors and of the
ideas discussed in this Letter. In addition, the first factor will be
tested by the future Probe Of Extreme Multi-Messenger Astrophysics
(POEMMA)~\cite{Olinto:2017xbi} and the Giant Radio Array for Neutrino
Detection (GRAND)~\cite{Alvarez-Muniz:2018bhp}, which may directly observe
neutrino-induced sphaleron transitions raining down
on the Earth atmosphere.\\

The research of LAA is supported by the by the U.S. National Science
Foundation (NSF Grant PHY-1620661) and the National Aeronautics and
Space Administration (NASA Grant 80NSSC18K0464). The research of I.A. is funded in
part by the ``Institute Lagrange de Paris", in part by the Swiss National Science Foundation 
and in part by a CNRS PICS grant.

\end{document}